# Fractal solutions of reactor models

## Marek Berezowski

Institute of Chemical Engineering and Physical Chemistry, Cracow University of Technology, ul. Warszawska 24, 31-155 Krakow, Poland

**Abstract**

Three kinds of fractal solutions of model of chemical reactors are presented. The first kind concerns the structure of Feigenbaum_s diagram on the limit of chaos. The second kind and the third one concern the effect of initial conditions on the dynamic solutions of models. In the course of computations two types of recirculation were considered, viz. the recirculation of mass (return of a part of products stream) and recirculation of heat (heat exchange in the external heat exchanger).

## 1. Introduction

As was shown in earlier papers, the solutions of models of chemical reactors may be of a very complex dynamic character, including chaos [1–5]. These solutions are illustrated in a clear way by steady-state diagrams of various type, a.o. by the so-called Feigenbaum diagram (Fig. 1). In the case when bifurcation parameter exceeds the value, which determines the entrance of the trajectory into the chaotic region, the system of diagram branches has a fractal structure at this place [6]. In the present work the method of determination of the fractal dimension of Feigenbaum tree, concerning a chemical reactors, is demonstrated. Also the effect of initial value of state variables of reactor, viz. degree of conversion $\alpha$ and temperature $\Theta$, upon the value and convergence of models solutions was investigated and coloured fractal images were obtained. For the sake of computations a.o. the Mandelbrot algorithm was applied.

## 2. Model

The mathematical model of a chemical reactor with recirculation of mass or heat comprises following balance equations:

mass

$$\frac{\partial \alpha}{\partial \tau} + \frac{\partial \alpha}{\partial \xi} = \phi(\alpha, \Theta) \tag{1}$$

heat

$$\frac{\partial \Theta}{\partial \tau} + \frac{\partial \Theta}{\partial \xi} = \phi(\alpha, \Theta) + \delta(\Theta_H - \Theta) \tag{2}$$

and algebraic balance equations, which constitute the boundary conditions resulting from the feedback:

$$\alpha(0, \tau) = f\alpha(1, \tau) \tag{3}$$
$$\Theta(0, \tau) = f\Theta(1, \tau) \tag{4}$$

The reaction kinetics may be described by Arrhenius-type relationship:

$$\iota(\alpha, \Theta) = (1 - fx)Da(1 - \alpha)^n \exp\left(\gamma \frac{\beta \Theta}{1 + \beta \Theta}\right) \tag{5}$$

In the case of recirculation of mass $x=1$ [1,3,4] and in the case of recirculation of heat $x=0$ [2,5].The numerical analysis presented in papers [1–5] has demonstrated that there exist the regions of parameters values of both reactor models, i.e. with mass recirculation and with thermal feedback, with oscillation-type solutions, chaos including, these regions being of considerable magnitude. The character of these solutions is well exposed by Feigenbaum tree in Fig. 1.

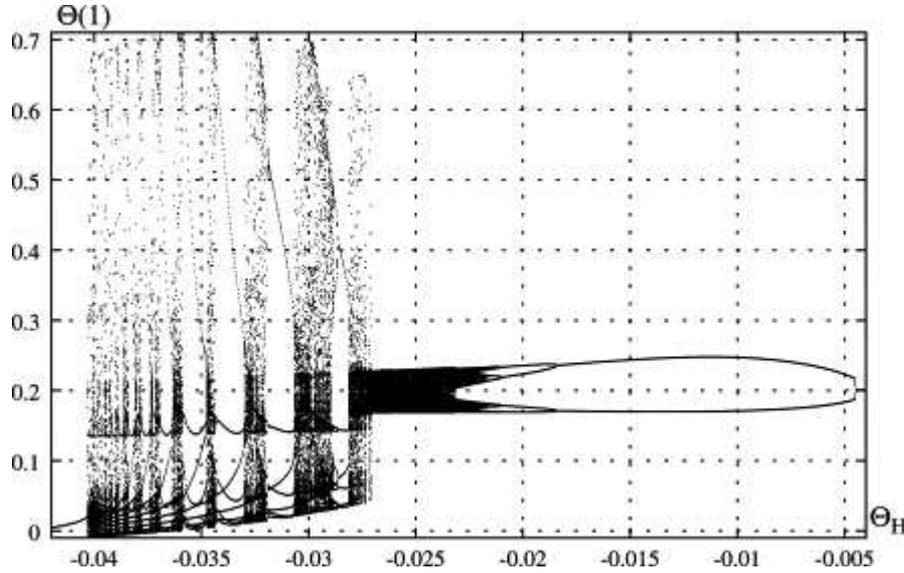

Fig. 1. Complete Feigenbaum's diagram of the reactor with recycle of mass. The effect of the temperature dependence.

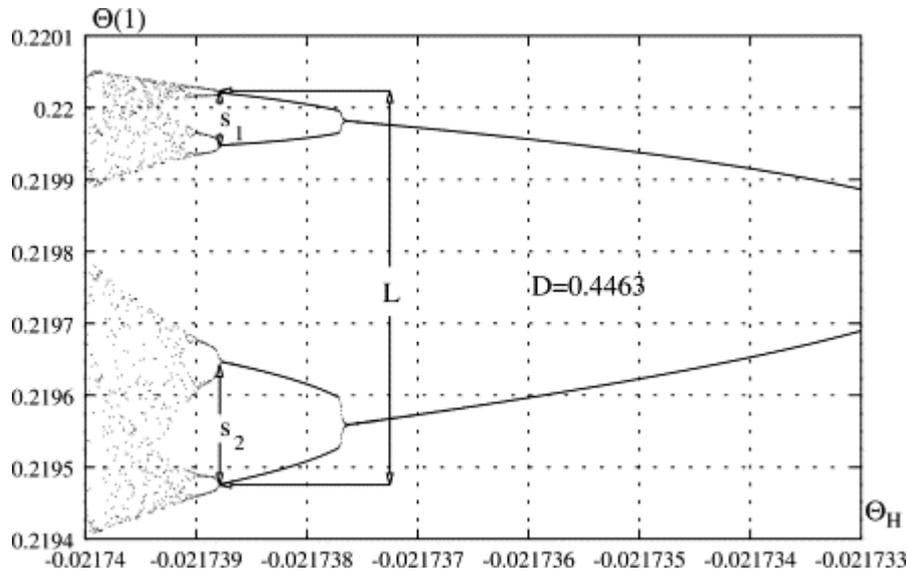

Fig. 2. Fragment of diagram from Fig. 1.

It concerns the reactor temperature $\Theta(1)$. For the computations the following values of parameters were assumed: Da=0.15, n=1.5, f=0.5, $\gamma=15$, $\beta=2$, $\delta=3$. It is likely to imagine that the structure of this tree, based on the scenario of doubling the period, resembles the constructions of Cantor set [6]. There is, however, a difference, because in Cantor set the ratio of segment division is strictly defined, being equal to $r=1/3$, whereas in the case of diagram from Fig. 1 this ratio is different for different bifurcation points. The fractal dimension $D$ of Cantor set, determined from Kolmogorov definition

$$2r^D = 2(1/3)^D = 1 \qquad (6)$$

equals $D = 0.630929753 \ldots$ In the case of Feigenbaum diagram from Fig. 1 the individual branches disperse according to two different division ratios $r_1$ and $r_2$, and not according to one only. It is easy to demonstrate (Appendix A) that Kolmogorov dimension $D$ may be, in this case, determined from the relationship

$$r_1^D + r_2^D = 1 \qquad (7)$$

This quantity should be determined at the Feigenbaum point, i.e. for the value of parameter $\Theta_H$, which—when exceeded—leads to the entry of trajectory into the chaotic region. In the analysed example, approximately, $\Theta_H = 0.021739$. In Fig. 2 the enlargement of a very small fragment of Fig. 1 is shown. Both division ratios were determined as

$$r_1 = \frac{s_1}{L}; \quad r_2 = \frac{s_2}{L} \qquad (8)$$

which, after applying Eq. (7), yields the fractal division $D=0.4463$. In order to check the correctness of the obtained result, the enlargement of a fragment of Feigenbaum diagram, this time referring to the reactor conversion degree $\alpha(1)$, is given in Fig. 3. It is clearly seen that the obtained leaves, after change of scale, are identical with the leaves in Fig. 2 (they overlap). The upper leaf from Fig. 2 is proportional to the lower one from Fig. 3, and the upper leaf from Fig. 3 is proportional to the lower one from Fig. 2. The change of place as above has no qualitative meaning and in each case the same value of dimension $D$ is obtained from Eq. (7).

## 3. Fractal images of the solutions

In this paper two types of coloured fractal images, obtained from the mathematical models of chemical reactors are presented. In order to determine them, two different algorithms were used.

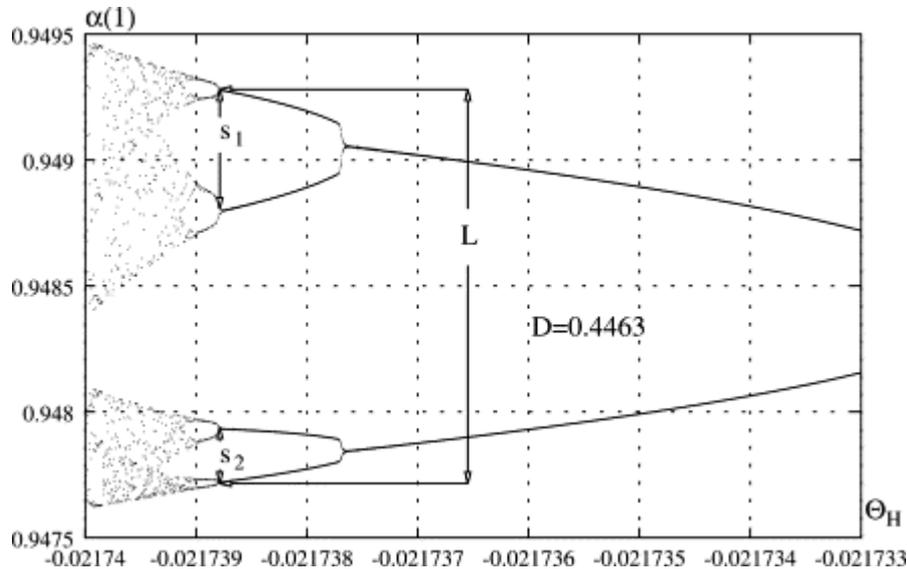

Fig. 3. Fragment of Feigenbaum diagram of the reactor with recycle of mass. The effect of degree of conversion.

### 3.1. Algorithm 1 of the generation of coloured fractal images

Generating a discrete time series of state variables of the reactor $\alpha(1.k)$ and $\Theta(1.k)$ [1,3–5] a finite number of recurrent computational steps N was assumed, after which points with the coordinates $\{\alpha(1,0), \Theta(1,0)\}$ were plotted on the diagram. The colour of plotted point depended on the interval, in which found itself the value of $\Theta(1,N)$, i.e. the value of temperature in the last, $N$th, iterative step. In such a way a coloured map of initial conditions, indicating their effect upon the value of solution, has been constructed. Assuming the values of parameters of reactor with recirculation of mass, as previously, and $\Theta_H = -0.023$, one has obtained image as in Fig. 4. The number of steps of the recurrence process N=200. It is likely to imagine that Fig. 4 resembles Saturn rings. It is worth mentioning that individual coloured bands in Fig. 4 display a clearly outlined granular structure. This part bears witness to the sensibility of solutions with respect to initial conditions. It is interesting that within a given band one of the colours is dominating.

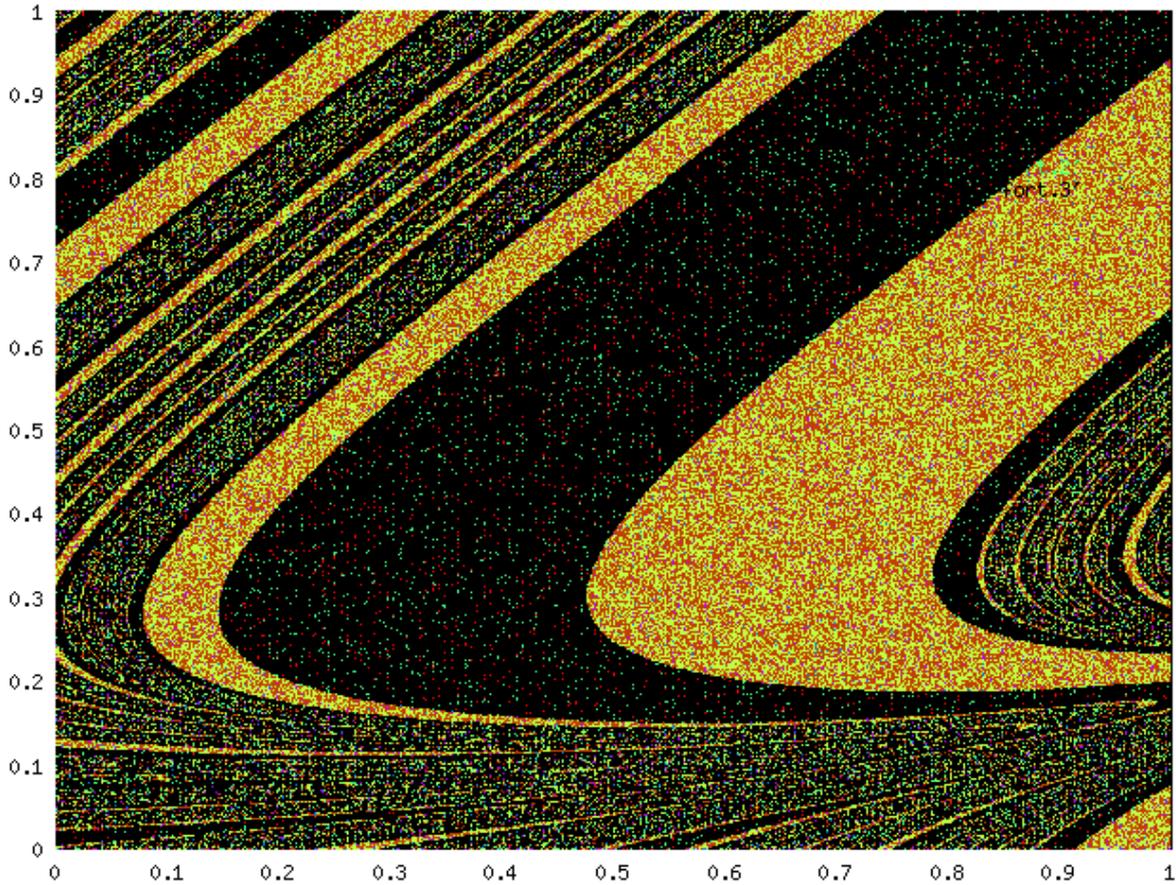

Fig. 4. Effect of initial conditions on the value of the solution of the model of reactor with the recycle of mass.

## 3.2. Algorithm 2 of the generation of coloured fractal images

In this case, for the generation of fractal image the method of Mandelbrot has been used, viz., it has been assumed that the state variables of reactor: conversion degree $\alpha$ and dimensionless temperature $\Theta$, are complex numbers. From the physical point of view this assumption is, obviously, unjustified. This results from the fact that physical variables are of physical character exclusively. Nonetheless it has been proved that the models of chemical reactors generate very interesting fractal structures for complex variables

$$\alpha = \alpha_r + i\alpha_i ; \quad \Theta = \Theta_r + i\Theta_i \qquad (9)$$

what is interesting from mathematical viewpoint.

The following algorithm has been assumed. For fixed values $\alpha(0,0)$ the map of initial conditions in the coordinate system $\{\Theta_r(1,0), \Theta_i(1,0)\}$ was constructed. The colour of a point on the map depended on the number of iterative steps k, after which the length of the vector $\left|\Theta\!\left(1,k\right)\right|$ exceeded the given value of $\varepsilon$. This means that the computations were terminated, when

$$\left|\Theta\!\left(1,k\right)\right| \geq \varepsilon \qquad\qquad\qquad (10)$$

If the absolute value $\left|\Theta\!\left(1,k\right)\right|$ exceeded $\varepsilon$ already in the first iterative step, no point was plotted on the diagram. In other words, the colours on the map correspond with the convergence rate of the numerical process. On the other hand, the points were coloured in red if the computational process was convergent, i.e. if after N steps the absolute value $\left|\Theta\!\left(1,N\right)\right|$ did not exceed $\varepsilon$. In such a way, changing successively $\Theta_r\!\left(1,0\right)$ and $\Theta_i\!\left(1,0\right)$ the fractal structures, as those on presented figures, were obtained.

So, Fig. 6 shows a basic fractal image of solutions of the model with thermal feedback (x=0).

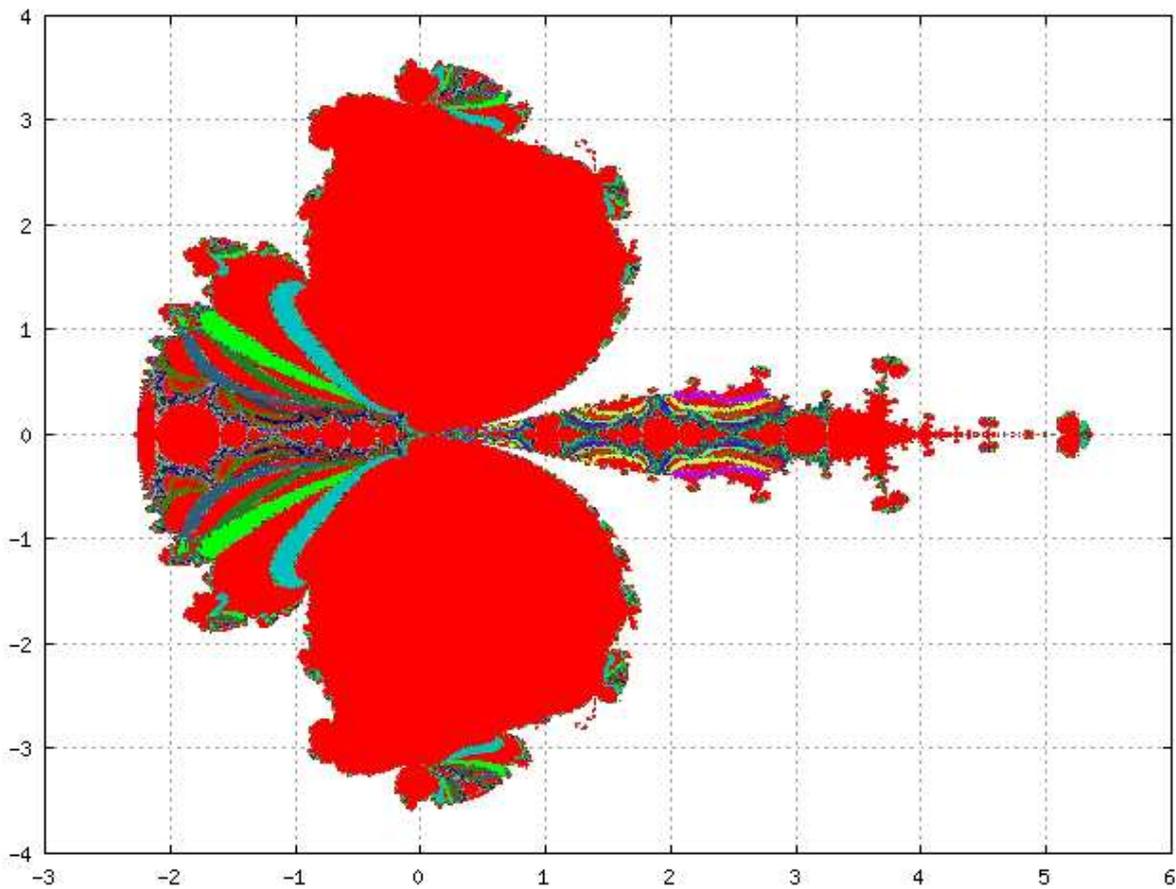

Fig. 6. Effect of initial conditions on the convergence of solutions of the model of reactor with the recycle of heat. General set.

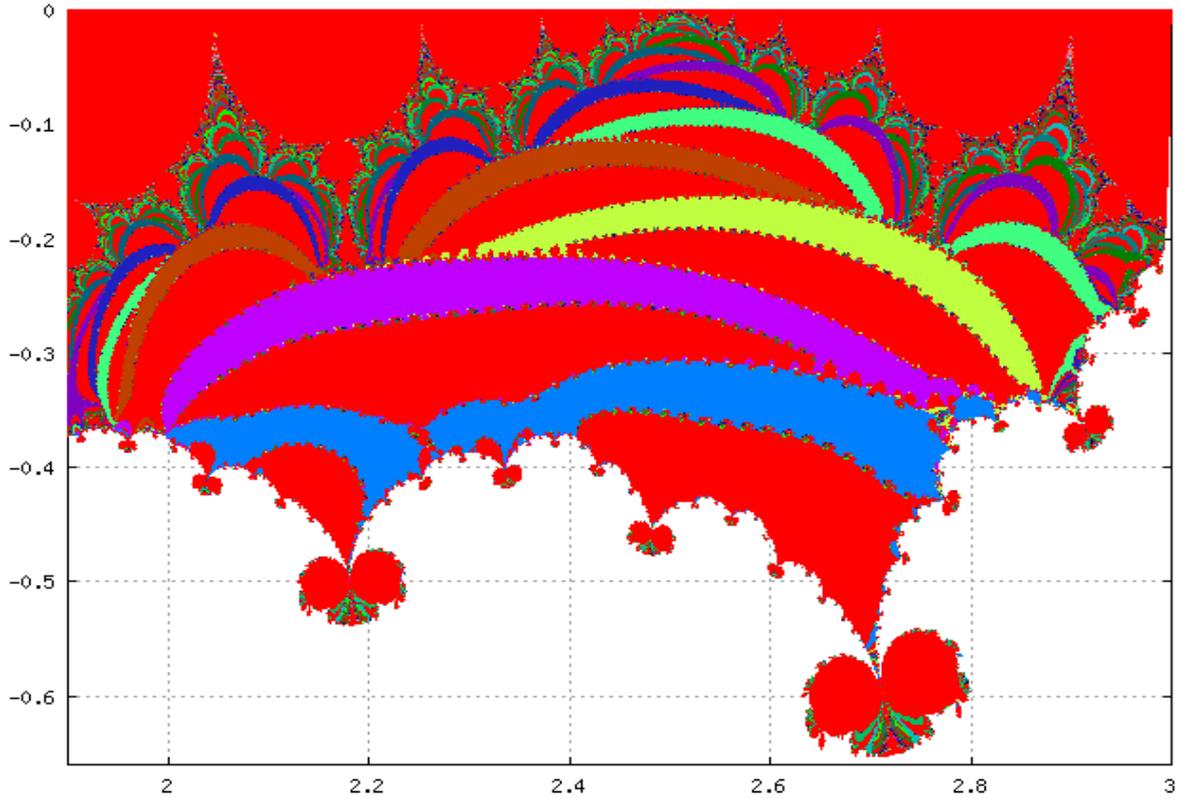

Fig. 7. Fragment of Fig. 6.

In this case the value of $\alpha(0,0)$ results from the nature of the process and is equal to $\alpha_r(0,0) = 0$, $\alpha_i(0,0) = 0$ [5]. For the sake of computations the following values of parameters: Da=0.15, n=1, f=0.3, $\gamma$=10, $\beta$=1.4, $\Theta_H$=-0:048, N= 200, $\varepsilon$=5 were assumed.

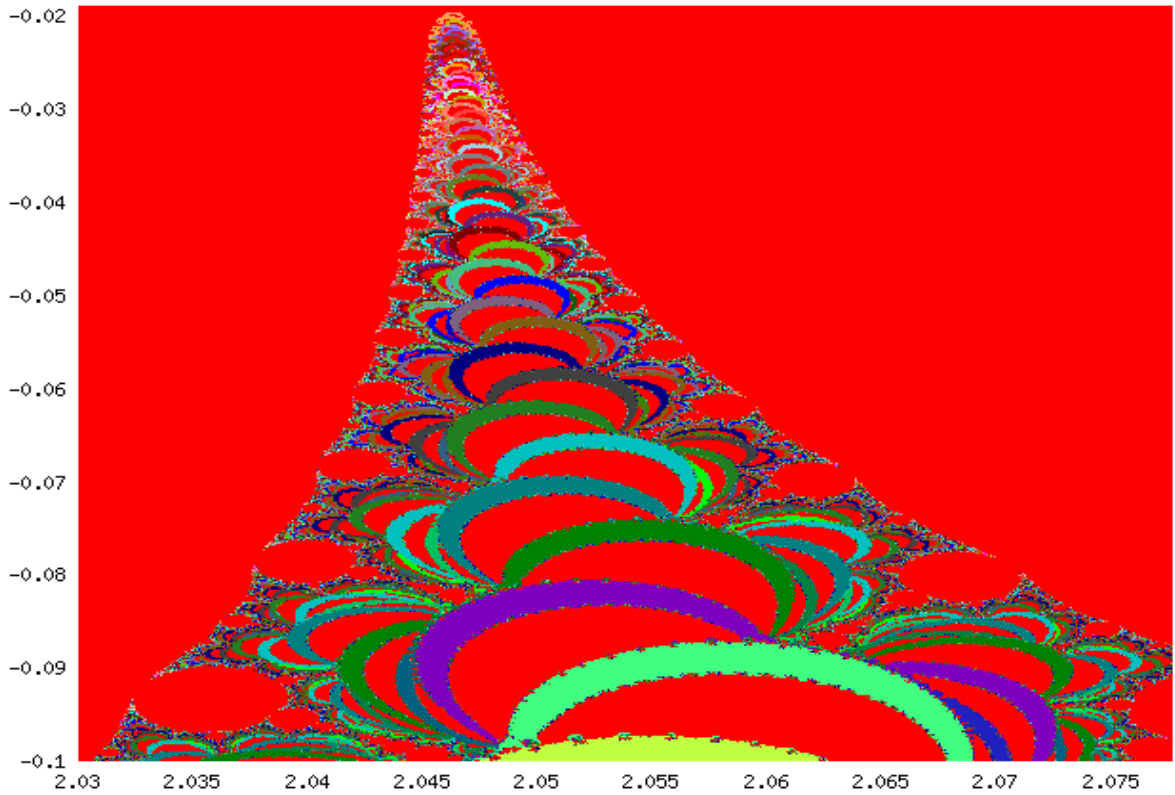

Fig. 8. Fragment of Fig. 7.

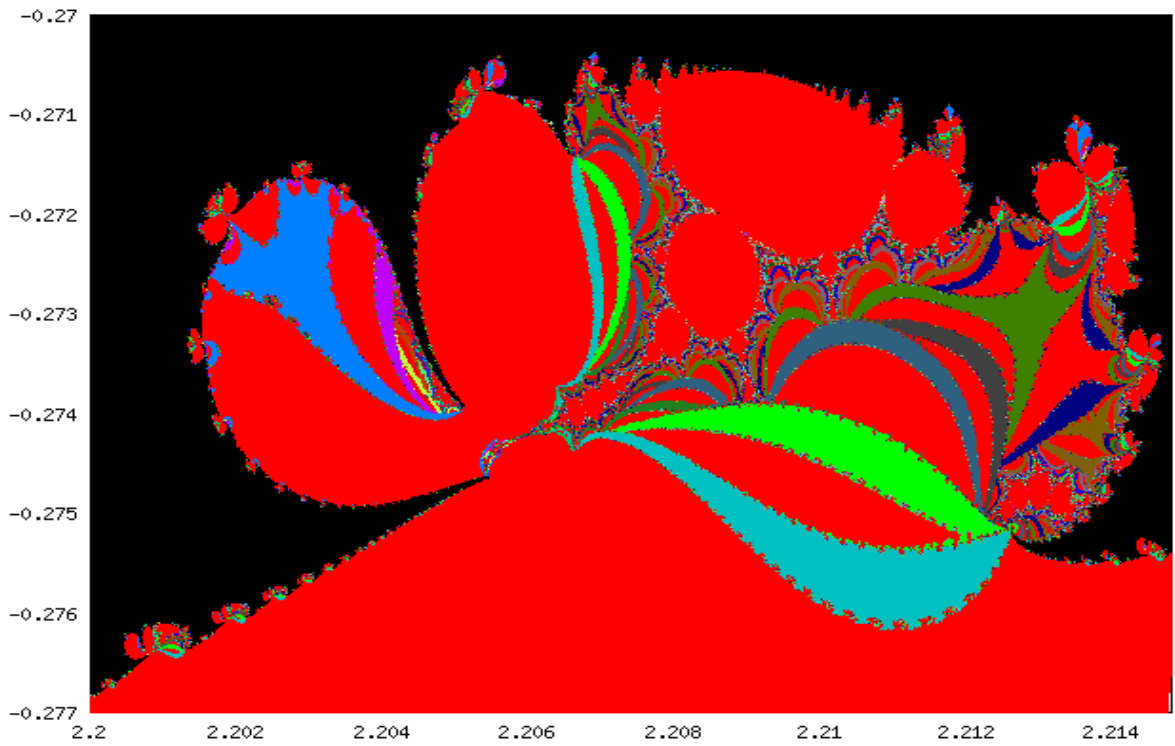

Fig. 9. Fragment of Fig. 7.

A fragment of the set from Fig. 6 is presented in Fig. 7. There are clearly visible regions resembling the form of a basic set, which bears witness to the fractal structure of the obtained solutions. The successive figures 8 and 9, present fragments of the set from Fig. 7. Also here the elements similar to the basic set are distinct.

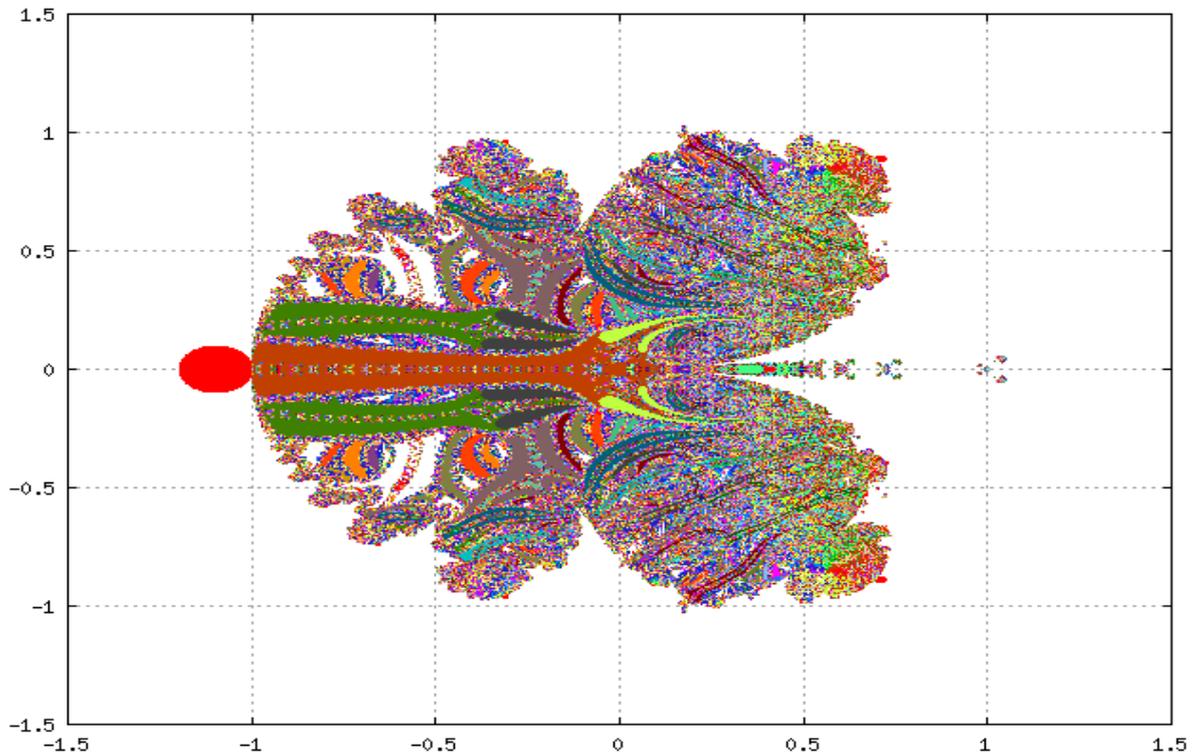

Fig. 10. Effect of initial conditions on the convergence of solutions of the model of reactor with the recycle of mass. General set.

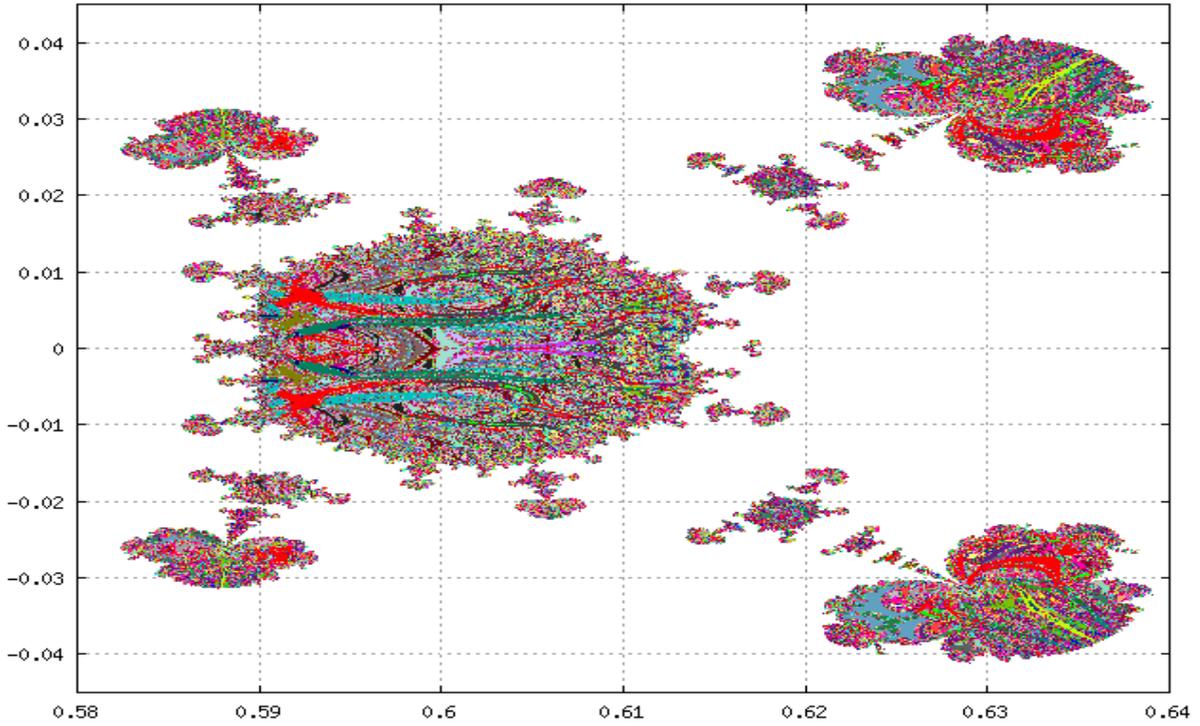

Fig. 11. Fragment of Fig. 10.

In Fig. 10 the basic fractal image of the solutions of the model of reactor with recirculation of mass (x=1) is shown [1,3,4]. In this case the values $\alpha_r(0,0)$=0.45, $\alpha_i(0,0)$=0, $\Theta_H$=-0.033, N=200 and $\varepsilon$=5 were assumed. The values of remaining parameters—as previously. The portions of the set from Fig. 10 are presented in Figs. 11 and 12. Similarly as in the case of recirculation of heat also here the forms of the basic set are clearly visible. Figs. 13 and 14 indicate the enlargement of the set from Fig. 11, whereas Fig. 15—the enlargement of the fragment of the set from Fig. 12.

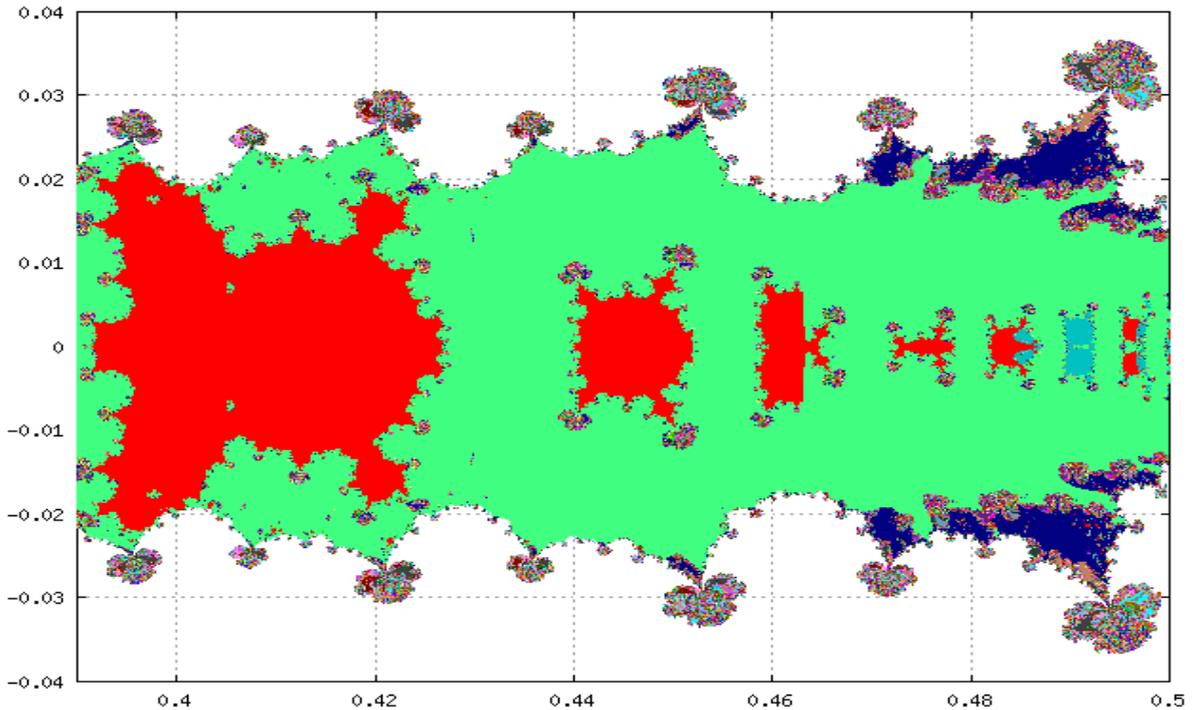

Fig. 12. Fragment of Fig. 10.

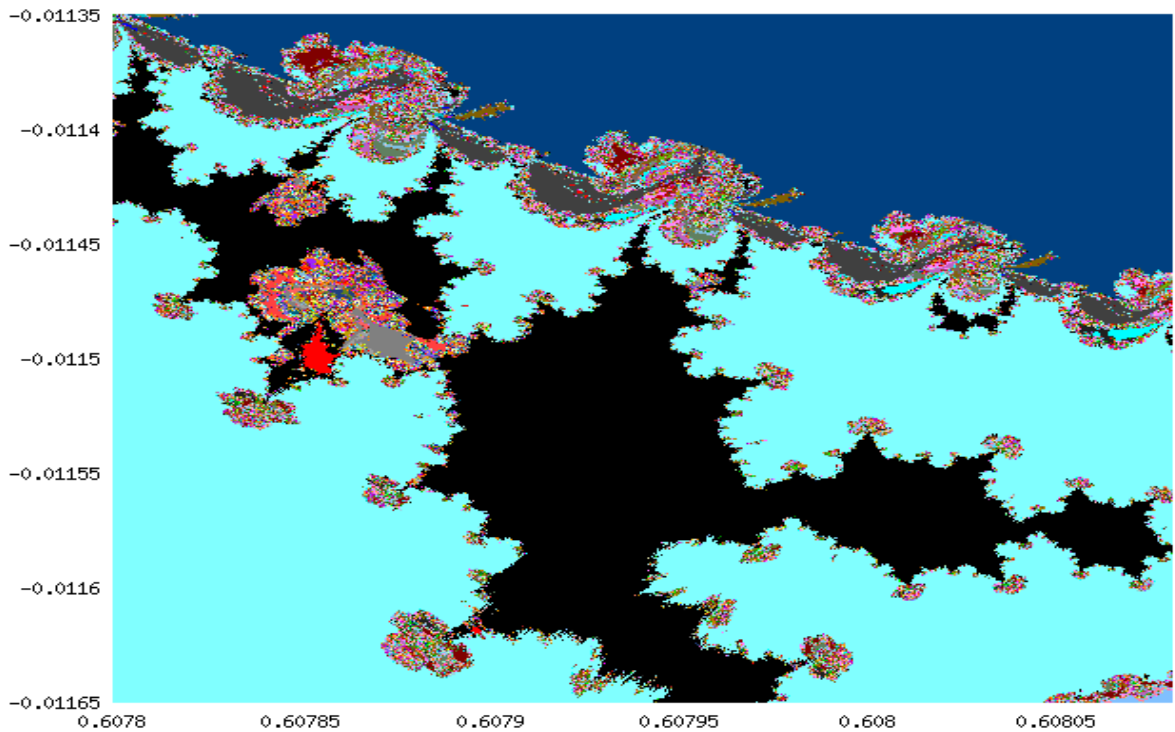

Fig. 13. Fragment of Fig. 11.

Comparing the images obtained from the model of reactor with thermal feedback with those obtained from model of reactor with recirculation of mass one observes that the structure of the former ones is more uniform, viz., the individual bands are of the same colour (compare Fig. 6). In the case of the reactor with recycle the colours are more strongly mixed (compare Fig. 10), which bears witness to a great turbulence of the process.

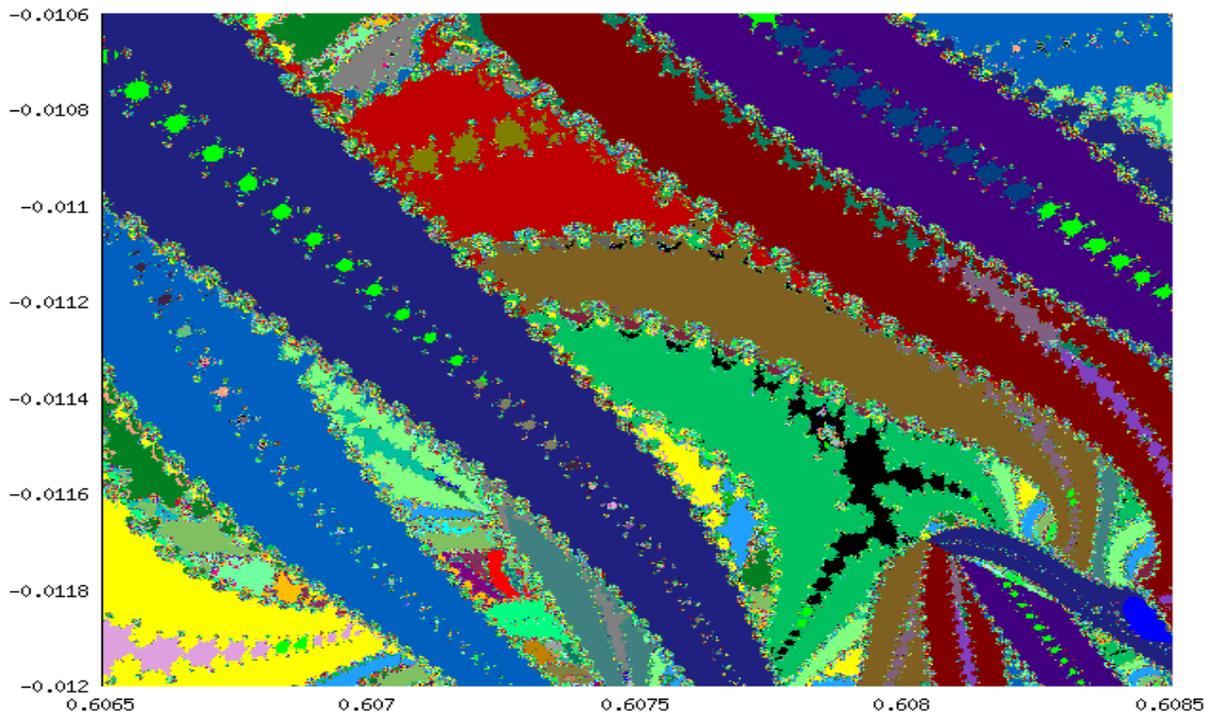

Fig. 14. Fragment of Fig. 11.

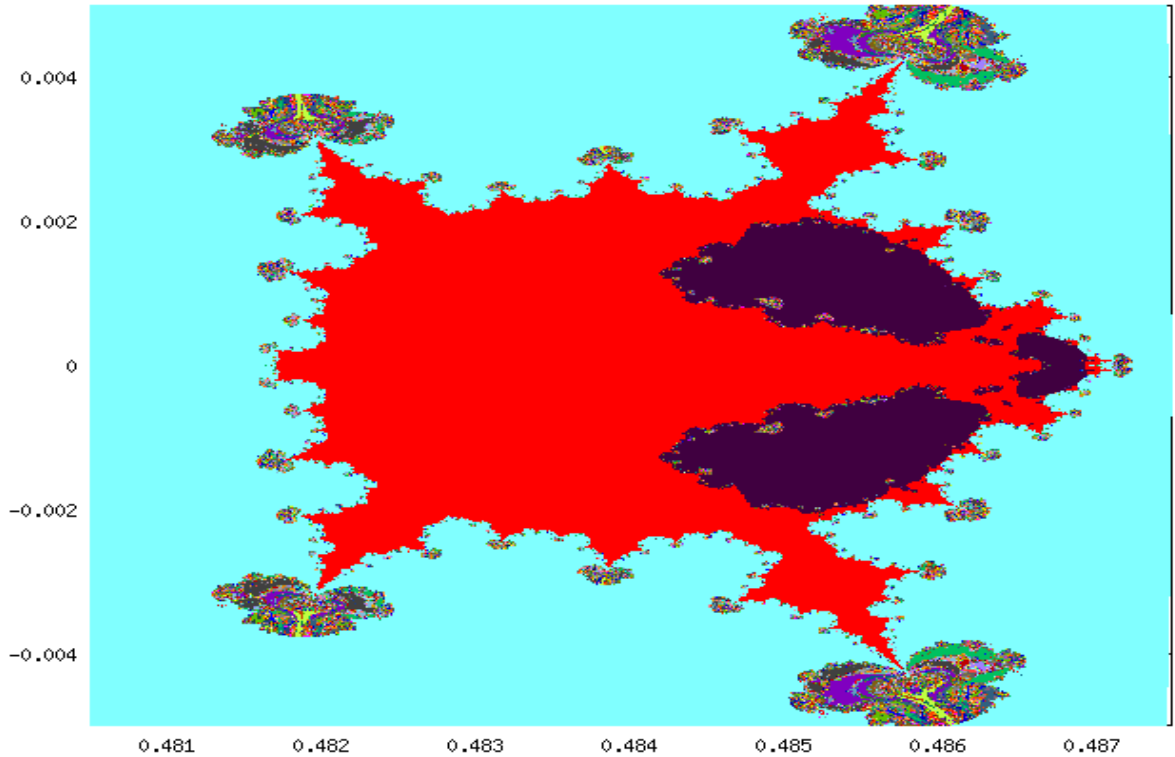

Fig. 15. Fragment of Fig. 12.

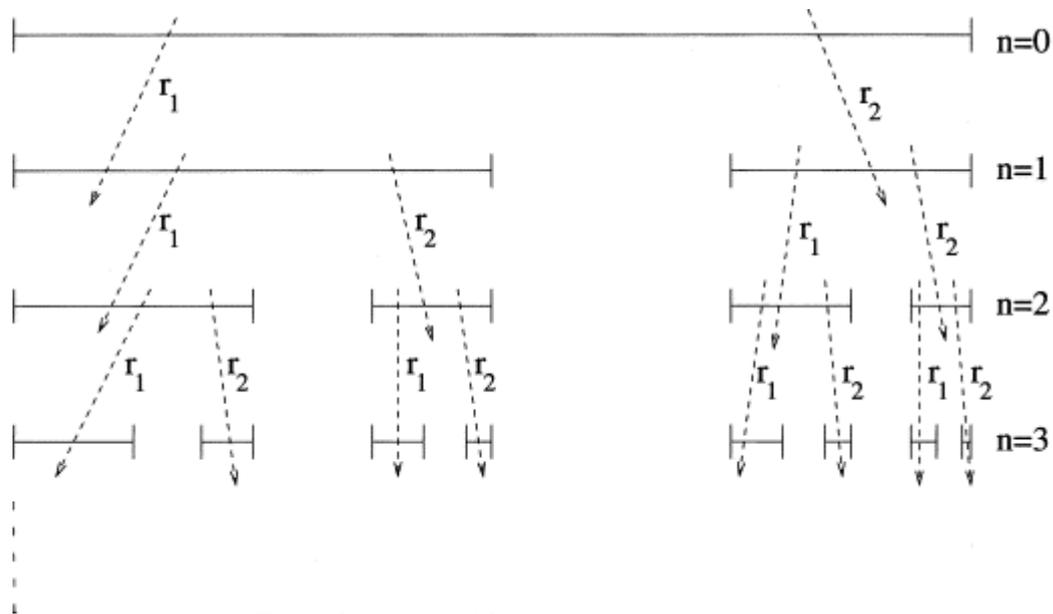

Fig. 16. Construction of Cantor_s set with two division ratios.

## 4. Concluding remarks

The present paper deals with three different models of fractal analysis of dynamic solutions of chemical reactor. The first fractal method concerns the analysis of Feigenbaum diagram of reactor model (Figs. 1–3). The branches of this diagram are treated as an analogue of Cantor set, the difference consisting in two division ratios in the case of reactor (Fig. 16). Finally, the formula (A.1), enabling one to determine the fractal dimension, has been derived. The second fractal method consists in the investigation of the effect of initial conditions of state variables, degree of conversion $\alpha$ and temperature $\Theta$ upon the terminal value (steady state) of these variables (Fig. 4). The third method, based on Mandelbrot algorithm, consists in the investigation of the convergence of solutions of the models, in dependence on initial conditions of these variables $\alpha$ and $\Theta$. It has been assumed, similarly as in Mandelbrot method, that these variables are of complex type. Finally, the maps of fractal images, as in Figs. 10–15, were obtained.

## Appendix A. Fractal dimension of a set with two different division ratios

Let us divide the closed segment [0,1] into three different parts, according to the division ratios $r_1$ and $r_2$ and remove the central part, leaving the boundary points. Let us do the same with two segments obtained $[0, r_1]$ and $[1 - r_2, 1]$, the same with four segments etc. (Fig. 16). In the limit a set analogous to Cantor set is obtained. Basing on Kolmogorov definition of the fractal dimension D, it is easy to notice that in the case of two different division ratios, $r_1$ and $r_2$, this dimension may be determined, in $n$th step of above algorithm, from Newton binomial

$$1 = \sum_{k=0}^{n} \binom{n}{k} \left(r_1^{n-k} r_2^k\right)^D = \left(r_1^D + r_2^D\right)^n \Rightarrow r_1^D + r_2^D = 1 \qquad (A1)$$

It follows from (A.1) that D, similarly as in case of Cantor set, does not depend on $n$, hence it is the general dimension of the set. The length of this set equals in the limit:

$$\lim_{n \to \infty} \left(r_1 + r_2\right)^n = 0 \qquad (A2)$$